# Gate-tunable exchange bias effect in FePS$_3$-Fe$_5$GeTe$_2$ van der Waals heterostructures

*Sultan Albarakati[1, 2#], Wen-Qiang Xie[3#], Cheng Tan[1#], Guolin Zheng[1*], Meri Algarni[1], Junbo Li [4], James Partridge[1], Michelle J. S. Spencer[1], Lawrence Farrar[1], Yimin Xiong[4], Mingliang Tian[4,5], Xiaolin Wang[6,7], Yu-Jun Zhao[3*], Lan Wang[1*]*

[1]ARC Centre of Excellence in Future Low-Energy Electronics Technologies (FLEET), school of Science, RMIT University, Melbourne, VIC 3001, Australia.

[2]Physics Department, Faculty of Science and Arts, P.O. Box 80200, Code 21589 Khulais, University of Jeddah, Saudi Arabia.

[3]Department of Physics, South China University of Technology, Guangzhou 510640, China.

[4]Anhui Province Key Laboratory of Condensed Matter Physics at Extreme Conditions, High Magnetic Field Laboratory, Chinese Academy of Sciences (CAS), Hefei 230031, Anhui, China.

[5]Department of Physics, School of Physics and Materials Science, Anhui University, Hefei 230601, Anhui, China.

[6]ARC Centre of Excellence in Future Low-Energy Electronics Technologies (FLEET), University of Wollongong, Wollongong, NSW 2500, Australia.

[7]Institute for Superconducting & Electronic Materials, Australian Institute of Innovative Materials, University of Wollongong, Wollongong, NSW 2500, Australia.

[#] These authors equally contributed to the paper. [*] Corresponding authors. Correspondence and requests for materials should be addressed to G. Z. (email: guolin.zheng@rmit.edu.au), Y.-J. Z. (email: zhaoyj@scut.edu.cn) and L. W. (email: lan.wang@rmit.edu.au).

# ABSTRACT

Electrical gate-manipulated exchange bias (EB) effect is a long-term goal for spintronics applications. Meanwhile, the emergence of van der Waals (vdW) magnetic heterostructures provides ideal platforms for the study of interlayer magnetic coupling. However, to date, the electrical gate-controlled EB effect has yet to be realized in vdW heterostructures. Here, for the first time, we realized electrically-controllable EB effects in a vdW antiferromagnetic (AFM)-ferromagnetic (FM) heterostructure, $FePS_3$-$Fe_5GeTe_2$. For pristine $FePS_3$-$Fe_5GeTe_2$ heterostructures, sizable EB effects can be generated due to the strong interface coupling, which also depend on the thickness of the ferromagnetic layers. By applying a solid protonic gate, the EB effects can be electrically tuned largely by proton intercalations and deintercalations. The EB field reaches up to 23% of the coercive field and the blocking temperature exceeds 50 K at $V_g = -3.15$ V. The proton intercalations not only tune the average magnetic exchange coupling, but also change the AFM configurations and transform the heterointerface between an uncompensated AFM-FM interface and a compensated AFM-FM interface. These alterations result in a dramatic modulation of the total interface exchange coupling and the resultant EB effects. The study is a significant step towards vdW heterostructure based magnetic logic for future low energy electronics.

**INTRODUCTION**

Exchange-bias (EB) effect, originating from the antiferromagnetic(AFM)-ferromagnetic (FM) interface coupling induced unidirectional anisotropy, has played a significant role in fundamental magnetics and spintronic device applications [1-6] since its discovery [7]. As a long time pursuit for spintronics applications, electrical gate-manipulated EB effects in AFM-FM structures enable scalable energy-efficient spin-orbit logic, which is very promising for beyond-COMS devices in future low energy electronic technologies [8]. To date, only very limited electrically tunable EB effects have been experimentally demonstrated, while most of them are based on oxide multiferroic thin film systems [9-12]. Recently, the emergence of van der Waals (vdW) heterostructures [13-15] and the discovery of 2D ferromagnetism [16, 17] have enabled various studies on vdW magnetic and spintronic devices [18-20]. High-quality interfaces and weak interlayer coupling in the vdW magnetic heterostructures suggest themselves as ideal platforms for exploring intrinsically interfacial magnetic coupling mechanisms, rather than the potential interfacial defects-dominated coupling in traditionally grown thin films [21-24]. So far, the EB effects have been extensively investigated in vdW magnets and AFM-FM heterostructures [25-30]. However, to date, an electrically tunable EB effect in vdW AFM-FM heterostructures has yet to be realized. Fortunately, a recent study [31] has demonstrated that solid protonic gates are effective tools for manipulating the interlayer magnetic coupling in single-crystalline $Fe_3GeTe_2$. Hence, applying solid protonic gates to vdW AFM-FM heterostructures may further unlock the door to the realization of electrically tunable EB in vdW AFM-FM heterostructures.

In this article, by utilizing a solid protonic gate, for the first time, we largely tuned the EB effects electrically in vdW AFM-FM heterostructures. Here, a $FePS_3$(AFM)-$Fe_5GeTe_2$(FM) vdW heterostructure is used as a model system which shows strong interface magnetic coupling at the AFM-FM interfaces. This coupling gives rise to a large EB effect that is influenced by the thickness of the Fe$_5$GeTe$_2$ (F5GT) layer and the amplitude of the cooling field $H_{cooling}$ but is insensitive to the thickness of the FePS$_3$ (FPS) layer ($\geq 15$ nm). Specifically, the EB effects are only observed in FPS-F5GT heterostructures within a small cooling field ($\leq 2$ T) and a narrow FM thickness range of 10 nm to 20 nm. Theoretical analysis based on density functional theory indicates that the proton intercalations affect the FPS-F5GT heterostructures in two significant ways. On the one hand, by intercalating the protons into the FPS-F5GT interface, the average interlayer exchange coupling $J_{ij}$ can be tuned; on the other hand, intercalation can also change the AFM configurations in the FPS layer and transform the FPS-F5GT heterointerface from an uncompensated AFM-FM interface to a compensated AFM-FM interface. This modulation of the AFM configurations results in a dramatic change of the total interface magnetic coupling $J_{int}$ and the resultant EB effects. This first realization of electrically controllable EB effect in a vdW AFM-FM heterostructure enables vdW heterostructure-based magnetic logic for future low energy electronics.

**RESULTS AND DISCUSSION**

**EB effects in the pristine FPS-F5GT vdW heterostructure**

In our AFM-FM heterostructures, the FM layer, F5GT, is a vdW ferromagnet with a Curie temperature $T_c \sim 300$ K in bulk [32]. The F5GT crystal is formed from thicker Fe−Ge slabs sandwiched by Te layers. The crystal structure of F5GT has a space group R3m with lattice parameters $a = 4.04$ Å and $c = 29.19$ Å. The AFM layer, FPS, has a monoclinic structure with the space group of C2/m. FPS is an Ising-type AFM with a Néel temperature $T_N \sim 123$ K [33, 34]. Figure 1a shows a schematic of our heterostructure device. Experimentally, a FPS layer was first stacked onto a F5GT layer to form an AFM-FM heterostructure. The heterostructure was then covered by a hBN layer and transferred onto the Pt contacts of thickness ~15 nm (see the inset of Fig. 1c, device #2). Figure 1b shows temperature dependent

EB effects in device #2 (with thicknesses $t_{F5GT} = 12$ nm, $t_{FPS} \sim 20$ nm) under cooling fields $H_{cooling} = \pm 1$ T. After field cooling from 150 K ($> T_N$), the hysteresis loops at low temperatures shift to the opposite directions of the cooling fields, thus producing negative EB effects (Fig. 1b). As temperature is increased, the EB fields ($H_{EB} = |H_{EB}^+ + H_{EB}^-|/2$) are gradually smeared out and are supressed above the blocking temperature $T_B = 20$ K due to thermal fluctuations, as shown in Fig. 1c. The emergence of the EB effect below 20 K in device #2 indicates a strong magnetic coupling at FPS-F5GT interface.

**Gate tunable EB effects in the FPS-F5GT vdW heterointerfaces**

In order to further explore the manipulation of EB effect, protons have been intercalated into the FPS-F5GT heterointerfaces using a protonic gate. Figure 2a shows a schematic of our gating device, where the FPS-F5GT bilayer is mounted on a solid protonic conductor with a gating electrode underneath. Hence, a solid proton field effect transistor (SP-FET) is formed. Figure 2b shows the optical image of our gating device #3, with an upper F5GT layer $t_{F5GT} = 16$ nm and lower FPS layer $t_{FPS} \sim 15$ nm. The heterostructure is covered entirely in-situ by a thin hBN layer, as shown in the atomic force microscope image in Fig. 2c. The gating voltages are applied between the bottom electrode (~10 nm Pt) and the source electrode. At low temperatures, device #3 exhibits robust EB effects irrespective of the gating voltage (Section 1.1, Supplementary Material). However, above 20 K, the EB effects demonstrate a strong dependence on the gate voltage. Figure 2d and Fig. 2e illustrate the gate-dependent EB effects in device #3 at T = 30 K and 40 K, respectively. At T = 30 K, device #3 exhibits large interface magnetic coupling with a distinct negative EB effect ($H_{EB} = 178$ Oe) at $V_g = 0$ V (here the EB effect is regarded as a "ON" state). Applying a negative voltage bias of −3.15 V, the EB effect becomes more prominent with a maximum EB field $H_{EB} = 310$ Oe, reaching 23% of the coercivity ($H_{EB}/H_C = 23\%$). However, the EB effect vanishes abruptly when the voltage bias is swept to $-3.64\ V$, indicating suppression of the interface magnetic coupling so that it is effectively turned "OFF". This absence of the EB effect within such a small gate voltage interval is indicative of the sensitivity of the interface magnetic coupling to proton intercalations or deintercalations. Figure 2f shows the EB amplitudes under different gate voltages at 30 K. Specifically, EB effects can be regarded as turning "ON" for $V_g \geq -3.15$ V ($|H_{EB}| > 80$ Oe), while it is effectively turned "OFF" for $V_g < -3.15$ V ($|H_{EB}| \leq 50$ Oe). Similar phenomena are also obtained at $T = 40\ K$. There is no EB effect at $V_g = 5.16$ V at T = 40 K, as shown in Fig. 2e, indicating negligible interface magnetic coupling. However, the EB effect re-emerges at $V_g = -3.15$ V with a magnitude of $H_{EB} = 151$ Oe, displaying an "ON" state of interface magnetic coupling. Sweeping the gate voltage again to $V_g = -4.82$ V, the EB effect vanishes again and hence an "OFF" state is realized. Figure 2g shows that consecutively sweeping the gate voltages from $-5.39$ V to 5.16 V causes alternate emergence ("ON") and suppression ("OFF") of the EB effects.

**Dependence of the EB effects on temperature, voltage, cooling field and thickness.**

The high tunability of the EB effects can be further verified when observing the gate dependent blocking temperature $T_B$. Specifically, no EB effect is detected with T > 50 K, indicating that $T_B$ is around 60 K at $-3.15$ V in device #3. However, $T_B$ drops down to 30 K at $-3.64$ V, as shown in Fig. 3a. By consecutively changing the voltages, we found $T_B$ varied between 30 K and 60 K, as illustrated in Fig. 3b. Compared to the low blocking temperature exhibited in device #2 (~20 K) without gating, the much higher blocking temperature in device #3 obtained at $-3.15$ V indicates a high tunability of the interlayer magnetic coupling under the protonic gating. Besides their strong dependence on gate voltage, the EB effects were also found to be substantially suppressed under large cooling fields, as shown in Fig. 3c. In device #3 with $T = 20$ K and $V_g = -4.13$ V, a small cooling field of $H_{cooling} = 1$ T generates a negative EB effect with an amplitude of about 126 Oe. However, when the cooling field is increased up to $H_{cooling} = 2$ T, the EB field decreases to around 60 Oe (shown as red, open squares). A similar trend is also observed in device #3 with $T = 30$ K and $V_g = 3.87$ V. For a small cooling field of $H_{cooling} = 0.6$ T, a large EB effect is observed with $H_{EB} \sim 190$ Oe. However, increasing the cooling field gradually decreases the EB effects. The EB effects are negligible when the cooling field $H_{cooling} > 2$ T (shown as blue open squares in Fig. 3c). Another interesting phenomenon is the dependence of EB effects on the thickness of FM layer, as shown in Fig. 3d. For these results, we maintained the FPS thickness $t_{FPS} > 15$ nm while varying the F5GT thickness $t_{F5GT}$. EB effects were observed in the FPS-F5GT heterostructures with $t_{F5GT} = 12$ nm (device #2), 16 nm (device #3), 18 nm (device #4). Among them, a maximum EB effect reaching $H_{EB} \sim 570$ Oe was detected in device #4 at 2 K with $V_g = 2.82$ V (Section 1.2, Supplementary Material). However, no EB effects were observed in device #1 ($t_{F5GT} = 5$ nm) and device #5 ($t_{F5GT} = 25$ nm), as shown in Fig. 3d (also in Section 1.2, Supplementary Material). Among them, devices #1, #3, #4 and #5 were mounted on protonic conductors while device #2 was on SiO$_2$ substrate without a protonic conductor. Generally, the EB effect is an interface effect which decreases as the thickness of the FM layer increases. This is consistent with experimental observations showing that no EB effects were obtained when $t_{F5GT} > 20$ nm. However, EB effects were not exhibited in heterostructure devices with thinner F5GT nanoflakes ($t_{F5GT} < 10$ nm), even under protonic gating. This is attributed to the extremely large coercivity ($H_c \sim 2$ T) characteristic of the thinner F5GT nanoflakes ($<$ 10 nm) arising from intralayer defect pinning [35]. The energy barrier induced by defect pinning is significantly larger than that from unidirectional anisotropy. Thus, the magnetic field required to overcome the pinning effect in the magnetization process would be sufficient to overwhelm the unidirectional anisotropy induced by the interface magnetic coupling and thereby suppress EB effects.

**Theoretical analysis based on density functional theory**

To gain more insight into the gate-dependent EB effects, we conducted a theoretical analysis using density functional theory (DFT) [36, 37]. The computational details are enclosed in section 2 of the Supplementary Material. In brief, the heterojunction models were constructed as follows. First, twelve different displaced combinations between FPS and F5GT were analysed (Fig. S6). The combination with the lowest adhesion energy (FPS/F5GT-up model) was then chosen for further calculations. Second, the adsorption sites of H in a single layer and heterojunction were optimized and the structure with the lowest adsorption energy (FPS/H/F5GT-up-1 model) was selected for analysis (Fig. S7 and Fig. S8). Finally, possible AFM magnetic configurations of the FPS layers were examined within our constructed heterointerfaces (Fig. S9).

Generally, the EB effect depends strongly on the interface coupling of AFM-FM system [38] and it is hard to generate using an Ising-type AFM because the interface is highly compensated [39]. The intensity of the EB effect relates to the difference of two neighbouring domains ($\Delta\sigma$) [39]. For an ideally uncompensated AFM-FM coupling system, $\sigma = \pm J_{int}/a_{lat}^2$, where $a_{lat}$ is the lattice parameter. $J_{int}$ is the total interface exchange coupling, and is equal to the integration of the average magnetic exchange coupling ($J_{ij}$) up to the given distance (Section 2.1 in Supplementary Material). For an ideally compensated system, $\sigma = J_{int} = 0$. In a F5GT/FPS heterostructure, due to the low $P1$ symmetry of the heterojunction, the interface is uncompensated, giving rise to a non-zero EB effect. The EB effect may be influenced by three factors, namely, the exchange coupling $J_{ij}$ at the heterointerface and the magnetic states of F5GT and FPS. According to previous experimental study [35], the protonic-gate cannot generate exchange bias in F5GT and nanoflakes with thicknesses $t_{F5GT} < 20$ nm also retain their ferromagnetic state under a protonic gate. Moreover, the coercivities of F5GT in our heterostructure devices are constant under gate voltage application. Hence, only the effect of interlayer coupling between FPS-F5GT and the AFM configurations in FPS are important for the explanation of our experiment results. We obtained two key results from our DFT calculations (Section 2.4-2.6 in Supplementary Material). First, the interlayer coupling at the FPS-F5GT interface decreases with increasing proton intercalation. As shown in Fig. 4a, intercalating H into the FPS-F5GT interface, the average magnetic exchange coupling $J_{ij}$ is decreased or reversed (Section 2.4 and 2.6 in Supplementary Material), resulting in a dramatic suppression of total interface magnetic coupling $J_{int}$. Second, FPS has several equivalent AFM configurations with small energy differences. Both protonic gate and magnetic field can induce a phase transition between an uncompensated AFM and a compensated AFM state (Section 2.5 in Supplementary Material). The three energy-favored equivalent configurations of the FPS layer FerriM_$2a$, FerriM_$2b$, and FerriM_$2c$ are shown in Figs 4b, 4c and 4d, respectively. As illustrated in Fig. 4e, the magnetic coupling energy $J_{int}^{FerriM\_2b}$ and $J_{int}^{FerriM\_2c}$ are significantly suppressed compared with $J_{int}^{FerriM\_2a}$, suggesting that the magnetic transformations between equivalent AFM configurations could also lead to a significant decrease of interface coupling (strong compensation).

Based on the two aforementioned key results, gate-dependent EB effects can be well explained. In Fig. 2, the protons can be intercalated into the FPS-F5GT interface at $V_g > 0$ V, which decreases the interface coupling. Hence, the 'OFF' state around $V_g = +5.16$ V at 40 K is due to the extremely small interface coupling after H insertion. Around 5.16 V, the thermal agitation energy at 40 K is larger than the AFM-FM coupling induced unidirectional anisotropy energy and hence EB disappears. EB values increase from $+5.16$ V to $-3.15$ V at 30 K and 40 K (also in Fig. S2) due to the deintercalation of the protons. The sharp transition from the

'ON" state to the "OFF' state at around −3.15 V can be attributed to the phase transition of FPS from an uncompensated AFM state (between +5.16 V and − 3.15 V ) to a compensated AFM state ( $V_g < -3.15$ V at 30 K and $-4.82$ V $< V_g < -3.15$ V at 40 K). Similar phase transition of FPS from a compensated AFM state to an uncompensated AFM state happens again at around −5.39V at 40 K, leading to an 'ON' state. Now we explain the variation of $T_B$ in Fig. 3a and Fig. 3b. At $V_g > -3.15$ V, $T_B$ is mainly determined by the interface exchange $J_{ij}$ of FPS-F5GT (See in Supplementary Material 2.4). Intercalating the protons into the interface decreases the interface coupling, resulting in a decrease of $T_B$ at $V_g > 0$ V. At $V_g < -3.15$ V, however, the transformation of FPS between an uncompensated AFM and a compensated AFM leads to a decrease (at − 4.82 V $< V_g < -3.15$ V) or an increase (at $-5.39$ V) of $T_B$. Note that, magnetic structure can also be tuned by the magnetic field [25, 40], which may lead to a magnetic transition from an uncompensated AFM to a compensated AFM in FPS. The compensated AFM is more stable in FPS at high magnetic fields, which explains the disappearance of the $E_B$ at 30 K when $|H_{cooling}| > 2$ T, as shown in Fig. 3f.

**CONCLUSION**

In conclusion, electrically tunable exchange bias effects in the vdW FPS-F5GT heterostructures has been reported, which originates from the proton intercalation tuned interface magnetic coupling in FPS-F5GT heterostructures. With the assistance of density functional theory, we found the interface exchange coupling is modulated either by direct intercalation of protons into FPS-F5GT interface or by transformations among equivalent AFM configurations in the FPS layer. Our study opens a route for vdW heterostructure-based magnetic logics for beyond-CMOS applications.

## METHODS

### Single Crystal Growth

Single crystal FPS and some F5GT single crystals were purchased from HQ graphene. Other F5GT crystals were grown by the transport method discussed in ref. 32.

### Device Fabrication and Transport Measurements

All hBN, F5GT and FPS flakes were mechanically exfoliated in a glove box with oxygen and water levels below 0.1 parts per million. We firstly used a large hBN layer to pick up the FPS layer and then the F5GT layer in-situ, the hBN/FPS/F5GT heterostructure was then transferred onto bottom Pt electrodes by a dry transfer technique. For the gating devices, we firstly used a thin hBN layer (thinner than $3\ nm$) to pick up the F5GT layer and then FPS layer to form hBN/F5GT/FPS heterostructures. These were finally transferred onto the solid proton conductors. Standard electron beam lithography was then used to fabricate the Hall bar devices. Before the Cr/Au evaporation, the top hBN layer in Hall bar area was etched by the Ar plasma for 90 s with an etching speed ~ $4\ nm/min$. The solid protonic electrolyte was prepared via the sol-gel process (described in ref. 18 in main text). Protonic gating experiments were performed in a commercial Magnetic Property Measurement System (MPMS) with a maximum magnetic field of $7\ T$. To decrease the leakage current, the gate voltage was swept at $250\ K$. Once the resistance change was observed, the device was rapidly cooled to low temperatures for magneto-transport measurements.

**ACKNOWLEDGEMENTS**

S.A, W.-Q. X. and C. T. contributed equally to this work. This research was performed in part at the RMIT Micro Nano Research Facility (MNRF) in the Victorian Node of the Australian National Fabrication Facility (ANFF) and the RMIT Microscopy and Microanalysis Facility (RMMF). Work at RMIT university was supported by the Australian Research Council Centre of Excellence in Future Low-Energy Electronics Technologies (Project No. CE170100039). Work at South China University of Technology was supported by the National Natural Science Foundation of China (Grant No. 12074126) and the Key Research and Development Project of Guangdong Province (Grant No. 2020B0303300001).

**Figure captions**

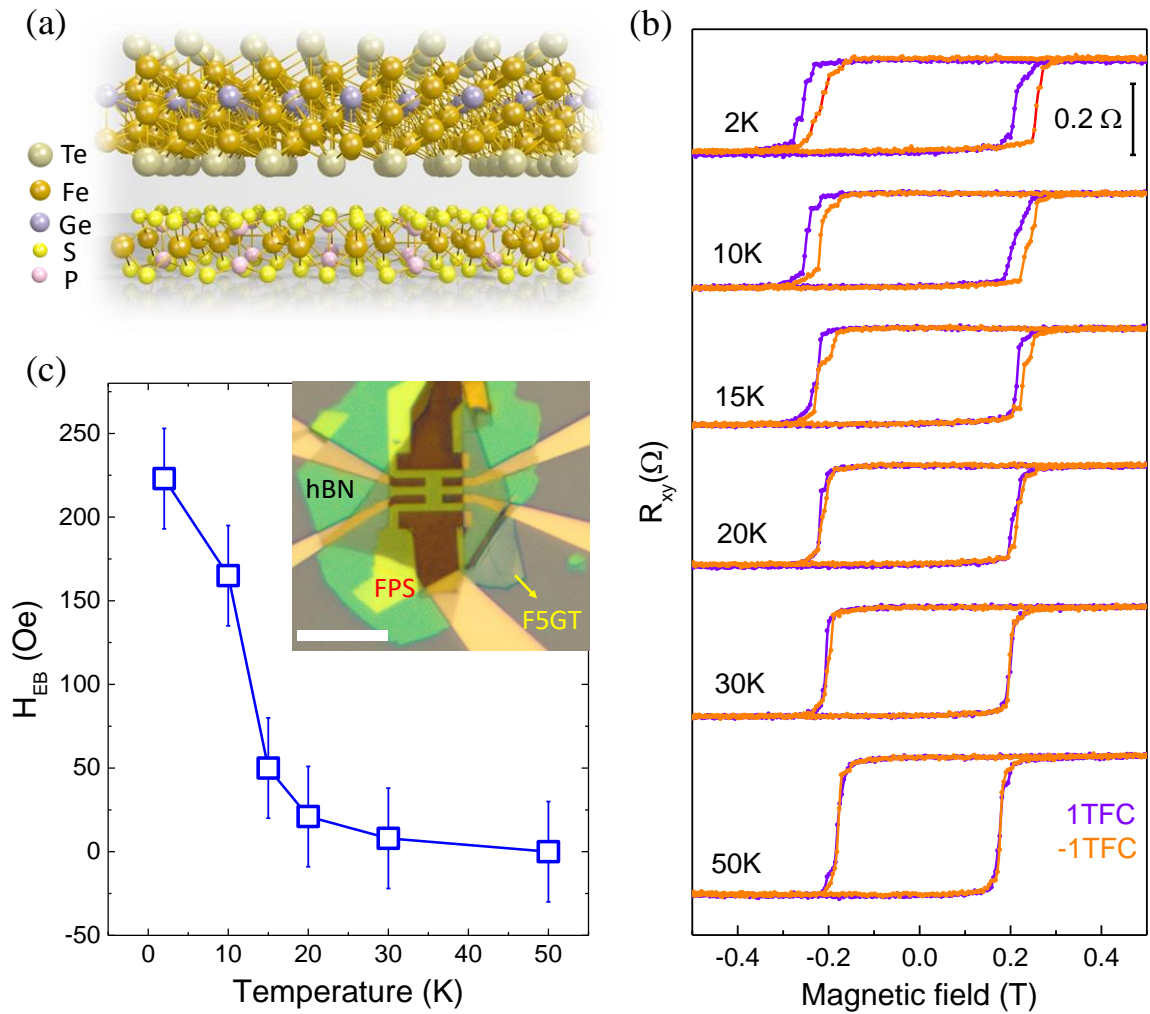

**Figure 1.** EB effects in a FPS-F5GT vdW heterostructure (device #2) without gating. (a) A schematic of the FPS-F5GT heterostructure. (b) Temperature-dependent EB effects in device #2 under cooling fields of $\pm 1\ T$. (c) The amplitudes of the EB effects at various temperatures. Inset: optical image of device #2. Scale bar: 10 $\mu m$. All loops are vertically shifted for clarity.

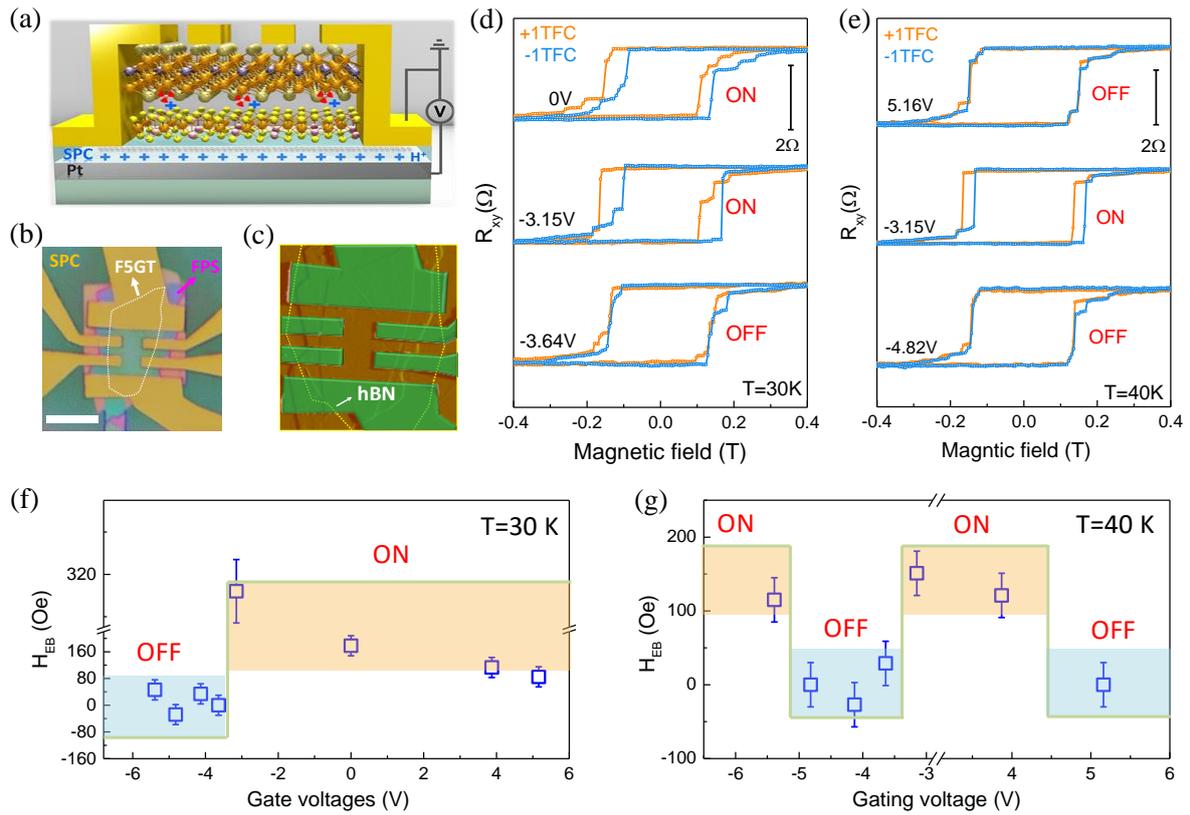

**Figure 2.** Gate tunable EB effects in the FPS-F5GT vdW heterointerfaces (device #3). (a) Schematic of the solid proton field effect transistor (SP-FET). (b, c) Optical and atomic force microscope- images of device #3. Scale bar: 10 $\mu m$. (d, e) Show the gate-dependent EB effects at $T = 30\ K$ and $40\ K$, respectively. (f, g) Illustrate the amplitudes of the EB effects under various gating voltages at $T = 30\ K$ and $40\ K$, respectively. All loops are vertically shifted for clarity.

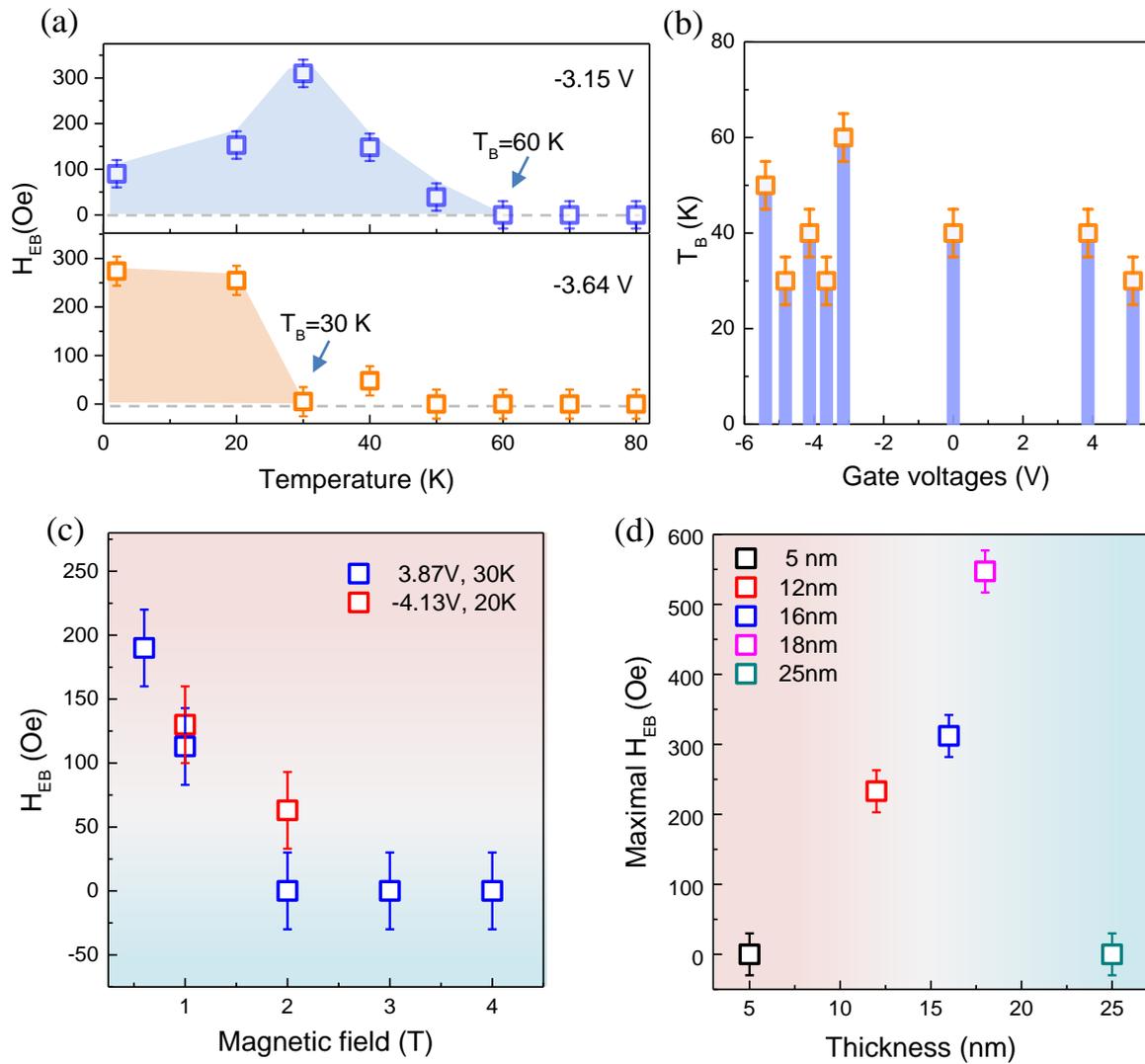

**Figure 3.** Dependence of the EB effects on temperature, voltage, cooling field and thickness. (a) Temperature dependent EB effects at selected gate voltages in device #3. (b) Gate dependent blocking temperature $T_B$ in device #3. (c) Cooling field dependent exchange bias effects in device #3. A large cooling field can significantly suppress the EB effects. (d) The thickness (F5GT) dependent maximum EB field. The EB effects only exist in F5GT with thickness between 10 $nm$ and 20 $nm$.

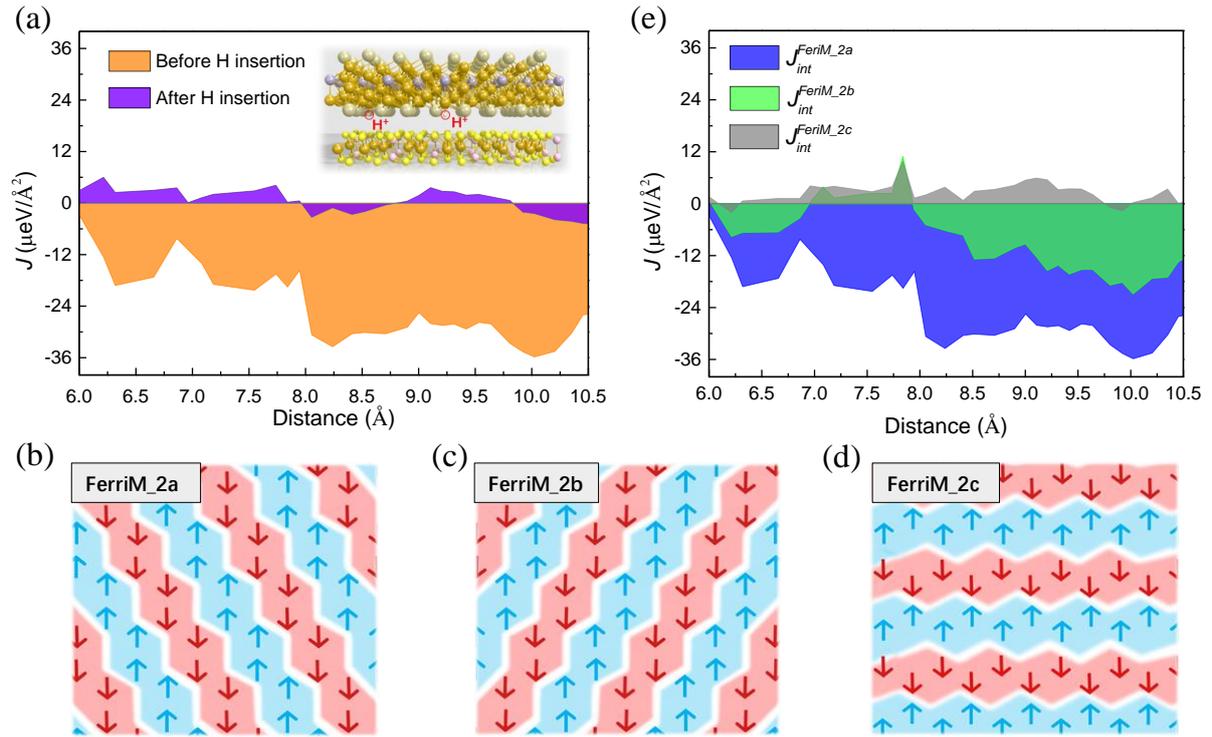

**Figure 4.** Theoretical analysis based on DFT. (a) The calculated interface magnetic couplings before (FPS/F5GT-up model) and after inserting H atom (FPS/H/F5GT-up-1 model) with a magnetic configuration of FerriM_2$a$. (b, c, d) Illustrate three equivalent magnetic configurations of the FPS layer in the interface. i.e., FerriM_2$a$, FerriM_2$b$, and FerriM_2$c$. $a$, $b$ and $c$ stand for three different zigzag orientations. The blue areas indicate "spin up" while the red ones suggest "spin down". (e) The calculated interface magnetic couplings of the FerriM_2$a$, FerriM_2$b$, and FerriM_2$c$ configurations with the FPS/F5GT-up model.